\documentclass[showpacs, showkeys, nofootinbib, floatfix]{revtex4}
\usepackage{amssymb}
\usepackage{amsmath}
\usepackage{graphicx}
\begin{document}

\title{Testing coupled dark energy with large scale structure observation}

\author{Weiqiang Yang$^a$}
\author{Lixin Xu$^{a,}$$^b$\footnote{lxxu@dlut.edu.cn}}

\affiliation{$^a$Institute of Theoretical Physics, School of Physics and Optoelectronic Technology, Dalian University of Technology, Dalian, 116024, P. R. China \\
$^b$State Key Laboratory of Theoretical Physics, Institute of Theoretical Physics, Chinese Academy of Sciences, Beijing, 100190, P. R. China
}

\begin{abstract}
The coupling between the dark components provides a new approach to mitigate the coincidence problem of cosmological standard model. In this paper, dark energy is treated as a fluid with a constant equation of state, whose coupling with dark matter is $\bar{Q}=3H\xi_x\bar{\rho}_x$. In the frame of dark energy, we derive the evolution equations for the density and velocity perturbations. According to the Markov Chain Monte Carlo method, we constrain the model by currently available cosmic observations which include cosmic microwave background radiation, baryon acoustic oscillation, type Ia supernovae, and $f\sigma_8(z)$ data points from redshift-space distortion. The results show the interaction rate in 3$\sigma$ regions: $\xi_x=0.00328_{-0.00328-0.00328-0.00328}^{+0.000736+0.00549+0.00816}$, which means that the recently cosmic observations favor a small interaction rate which is up to the order of $10^{-2}$, meanwhile, the measurement of redshift-space distortion could rule out the large interaction rate in the 1$\sigma$ region.
\end{abstract}

\pacs{98.80.-k, 98.80.Es}
\maketitle

\section{Introduction}

In the theoretical frame of general relativity, the accelerated expansion of the Universe could be successfully explained by so-called dark energy: an exotic and mysterious component with negative pressure. Recently, the Planck data \cite{ref:Planck2013,ref:Planck2013-CMB,ref:Planck2013-params}, covering a wide range of angular scales up to multipole number $l\sim2500$ with respect to the measurements of the cosmic microwave background (CMB) anisotropy, tells us that dark energy occupies about $68\%$ of the Universe, dark matter accounts for $28\%$, and baryonic matter occupies $4\%$. The simplest dark energy model consists of the cosmological constant and cold dark matter, which is dubbed as the $\Lambda$CDM model. Although the standard scenario is in good agreement with the new observational data, it meets with a fundamental puzzle, that is, the coincidence problem \cite{ref:Zlatev1999,ref:Chimento2003,ref:Huey2006}: why the densities of dark matter and dark energy are of the same order of magnitude, given that they evolve so differently with redshift?
%The wCDM model, which is composed by cold dark matter and dark energy with a constant equation of state (EoS), could be a possible way to keep away from this issue.

An effective method to alleviate the coincidence problem is to consider the coupling between dark matter and dark energy. The coupling allows the energy and momentum exchanges in the dark components, it would be possible to evoke such coupled mechanism to help mitigate the coincidence problem. This is a primary motivation for coupled dark energy. Since the nature of both dark matter and dark energy remain a mystery, there is as yet no reliable basis in fundamental theory for determining the coupling form. In most cases, the coupled dark energy models are on the basis of phenomenological consideration \cite{
% 2rd interaction H-Q
ref:Potter2011,ref:Aviles2011,
ref:Caldera-Cabral2009,ref:Caldera-Cabral2009-2,ref:Boehmer2010,ref:Boehmer2008,ref:Song2009,ref:Koyama2009,
ref:Majerotto2010,ref:Valiviita2010,ref:Valiviita2008,ref:Jackson2009,ref:Clemson2012,
ref:Bean2008,ref:Bean2008-2, % the two are no marks
% 3rd interaction phi-Q
ref:Chongchitnan2009,ref:Gavela2009,ref:Gavela2010,ref:Yang2014-uc,ref:Salvatelli2013,ref:Quartin2008,
ref:Honorez2010,ref:Costa2013,ref:Bernardis2011,ref:He2011,ref:He2010,
ref:He2008,ref:Abdalla2009,ref:Sadjadi2010,ref:Olivares2008,ref:Olivares2006,
ref:Olivares2005,ref:Sun2013,ref:Sadjadi2006,ref:Sadeghi2013,ref:Zhang2013,ref:Koivisto2005,ref:Simpson2011,
ref:Bertolami2007,ref:Avelino2012,ref:Quercellini2008,ref:Mohammadi2012,ref:Sharif2012,ref:Wu2007,
ref:Barrow2006,ref:Zimdahl2001,
% 3rd m1 interaction
ref:Lip2011d,ref:Chen2011,ref:Chen2009,ref:Koshelev2009,ref:Zhang2012,ref:Cao2011,ref:Guo2007,
% 3rd m2 interaction
ref:Liyh2013,ref:Bolotin2013,ref:Chimento2013,ref:Chimento2012RDE,ref:Chimento2011RDE}. A kind of popular method to specialize the coupling form is designing it as the linear combinations about the energy densities of the dark components
\begin{eqnarray}
\bar{Q}=A_c\bar{\rho}_c+A_x\bar{\rho}_x,
\label{eq:Qform}
\end{eqnarray}
where $\bar{Q}$ denotes the background energy exchange between dark matter and dark energy, $A_i=\Gamma_i$ or $A_i=3\xi_iH$ ($i=c,x$), both $\Gamma_i/H_0$ and $\xi_i$ are constants which might be called as interaction rate, whose absolute values of ranges are generally $[0,1]$ in the cosmological constraints \cite{ref:Clemson2012,ref:Salvatelli2013,ref:Costa2013,ref:Liyh2013,ref:Chimento2013}. The first type of coupled model is only related to the energy densities of the dark fluids, which is determined by local quantities \cite{ref:Potter2011,ref:Aviles2011,
ref:Caldera-Cabral2009,ref:Caldera-Cabral2009-2,ref:Boehmer2010,ref:Boehmer2008,ref:Song2009,ref:Koyama2009,
ref:Majerotto2010,ref:Valiviita2010,ref:Valiviita2008,ref:Jackson2009,ref:Clemson2012}. The second type of coupled model is proportional to the Hubble parameter and the energy densities of the dark components, which is influenced by the expansion rate of the Universe \cite{ref:Chongchitnan2009,ref:Gavela2009,ref:Gavela2010,ref:Yang2014-uc,ref:Salvatelli2013,
ref:Quartin2008,ref:Honorez2010,ref:Costa2013,ref:Bernardis2011,ref:He2011,ref:He2010,
ref:He2008,ref:Abdalla2009,ref:Sadjadi2010,ref:Olivares2008,ref:Olivares2006,
ref:Olivares2005,ref:Sun2013,ref:Sadjadi2006,ref:Sadeghi2013,ref:Zhang2013,ref:Koivisto2005,ref:Simpson2011,
ref:Bertolami2007,ref:Avelino2012,ref:Quercellini2008,ref:Mohammadi2012,ref:Sharif2012,ref:Wu2007,
ref:Barrow2006,ref:Zimdahl2001}. In addition, some other coupling forms have been discussed in Refs.
%%%%%%% modify01 the third interaction (have H) (5+ papers)------ product, power of \rho_c and \rho_x
\cite{ref:Lip2011d,ref:Chen2011,ref:Chen2009,ref:Koshelev2009,ref:Zhang2012,ref:Cao2011,ref:Guo2007,
%%%%%%% modify02 the third interaction (have H) (5+ papers)------ dot of \rho_c or \rho_x
ref:Liyh2013,ref:Bolotin2013,ref:Chimento2013,ref:Chimento2012RDE,ref:Chimento2011RDE,ref:Forte2013}. According to the phenomenological approach, in a flat Friedman-Robertson-Walker (FRW) universe, the coupling could be introduced into the background conservation equations for dark matter and dark energy
\begin{eqnarray}
\bar{\rho}'_c+3\mathcal{H}\bar{\rho}_c=-a\bar{Q},
\label{eq:rhoc}
\end{eqnarray}
\begin{eqnarray}
\bar{\rho}'_x+3\mathcal{H}(1+w_x)\bar{\rho}_x=a\bar{Q},
\label{eq:rhox}
\end{eqnarray}
where the subscript $c$ and $x$ respectively stand for dark matter and dark energy, the prime denotes the derivative with respect to conformal time $\tau$, $a$ is the scale factor of the Universe, $\mathcal{H}=a'/a$ is the conformal Hubble parameter, $w_x=\bar{p}_x/\bar{\rho}_x$ is the equation of state (EoS) parameter of dark energy. $\bar{Q}>0$ presents that the direction of energy transfer is from dark matter to dark energy, which changes the dark matter and dark energy redshift dependence acting as an extra contribution to their effective EoS; $\bar{Q}<0$ means the opposite direction of the energy exchange.

The coupling between the dark components could significantly affect not only the expansion history of the Universe, but also the evolutions of the density perturbations, which would change the growth history of cosmic structure, one can see Refs. \cite{ref:Clemson2012,ref:Caldera-Cabral2009,ref:Song2009,ref:Koyama2009,ref:Honorez2010,ref:Koshelev2009}. Moreover, the structure growth of dark matter could distinguish the coupled dark energy model from the uncoupled one even if the two models own the almost identical background evolutions. In other words, the large scale structure information could provide the key to break the possible existing degeneracy at the background level. In the linear regime, peculiar velocity makes the galaxy redshift-space power spectrum anisotropic when the underlying real space linear power spectrum is isotropic. This anisotropy is dubbed as redshift-space distortion (RSD), which allows to measure the amplitude of fluctuations in the velocity field. Based on the RSD test, a model-dependent measurement of $f\sigma_8(z)$ has been suggested in Ref. \cite{ref:fsigma8-DE-Song2009}, where $\sigma_8(z)$ is the overall normalisation of the matter density fluctuations. The RSD measurements have been released from a variety of galaxy surveys in Table \ref{tab:fsigma8data}, which include the 2dFGRS \cite{ref:fsigma81-Percival2004}, the WiggleZ \cite{ref:fsigma82-Blake2011}, the SDSS LRG \cite{ref:fsigma83-Samushia2012}, the BOSS CMASS \cite{ref:fsigma84-Reid2012}, and the 6dFGRS \cite{ref:fsigma85-Beutler2012} surveys. These measurements from large scale structure could been used to carry out the constraints on the cosmological models. Under the inspiration of this idea, in combination with the RSD test and geometric measurements which mainly include CMB, baryon acoustic oscillation (BAO), and type Ia supernovae (SNIa), cosmological constraints on several models have been investigated in Refs. \cite{ref:fsigma8-HDE-Xu2013,ref:fsigma8andPlanck-MG-Xu2013,ref:Xu2013-DGP,ref:Yang2013-wdm,ref:Yang2013-cass,
ref:fsigma8total-Samushia2013}, the results show that information provided by the RSD-derived $f\sigma_8(z)$ test significantly enhances the precision of the constraints on the cosmological parameters compared to the case where only geometric measurements are adopted. In particular, for the coupled dark energy model, the geometry measurements mildly favor the coupling between the dark components \cite{ref:Clemson2012,ref:Salvatelli2013,ref:Costa2013,ref:Liyh2013,ref:Chimento2013}, at the same time, the measurement about the growth rate of dark matter perturbations possibly rules out large interaction rate, which was indicated in Ref. \cite{ref:Clemson2012}. Therefore, the $f\sigma_8(z)$ test would provide a very powerful and robust constraints on the interaction rate. Along this point of view, Yang and Xu combined the geometric tests with the RSD measurement to constrain the dark energy model when the momentum transfer was vanished in the dark matter frame \cite{ref:Yang2014-uc}, the joint data sets are able to estimate the parameter space to high precision and evidently tighten the constraints. They concluded that the recently cosmic observations favored a small interaction rate between the dark components, at the same time, the measurement of redshift-space distortion could rule out the large interaction rate in 1$\sigma$ region. In this paper, we also try to add the RSD measurement to constrain the coupled dark energy model for the disappeared momentum transfer in the rest frame of dark energy. It is worthwhile to anticipate that the large scale structure measurement will help to significantly tighten the cosmological constraints and some evidences could rule out large interaction rate.

\begin{table}
\begin{center}
\begin{tabular}{ccc}
\hline\hline $z$ & $f\sigma_8(z)$ & survey and references \\ \hline
$0.067$ & $0.42\pm0.06$ & $6dFGRS~(2012)$ \cite{ref:fsigma85-Beutler2012}\\
$0.17$ & $0.51\pm0.06$ & $2dFGRS~(2004)$ \cite{ref:fsigma81-Percival2004}\\
$0.22$ & $0.42\pm0.07$ & $WiggleZ~(2011)$ \cite{ref:fsigma82-Blake2011}\\
$0.25$ & $0.39\pm0.05$ & $SDSS~LRG~(2011)$ \cite{ref:fsigma83-Samushia2012}\\
$0.37$ & $0.43\pm0.04$ & $SDSS~LRG~(2011)$ \cite{ref:fsigma83-Samushia2012}\\
$0.41$ & $0.45\pm0.04$ & $WiggleZ~(2011)$ \cite{ref:fsigma82-Blake2011}\\
$0.57$ & $0.43\pm0.03$ & $BOSS~CMASS~(2012)$ \cite{ref:fsigma84-Reid2012}\\
$0.60$ & $0.43\pm0.04$ & $WiggleZ~(2011)$ \cite{ref:fsigma82-Blake2011}\\
$0.78$ & $0.38\pm0.04$ & $WiggleZ~(2011)$ \cite{ref:fsigma82-Blake2011}\\
$0.80$ & $0.47\pm0.08$ & $VIPERS~(2013)$ \cite{ref:fsigma86-Torre2013}\\
\hline\hline
\end{tabular}
\caption{The data points of $f\sigma_8(z)$ measured from RSD with the survey references. The former nine data points at $z\in[0.067,0.78]$ were summarized in Ref. \cite{ref:fsigma8total-Samushia2013}. The data point at $z=0.8$ was released by the VIPERS in Ref. \cite{ref:fsigma86-Torre2013}. Then, a lower growth rate from RSD than expected from Planck was also pointed out in Ref. \cite{ref:fsigma87-Macaulay2013}.}
\label{tab:fsigma8data}
\end{center}
\end{table}

This paper is organized as follows. In the next section, for the $\xi$wCDM model (the coupled wCDM model, which is composed by cold dark matter and dark energy with a constant EoS), we review the perturbation equations of the dark components in the rest frame of dark energy. Moreover, we consider $\bar{Q}=3H\xi_x\bar{\rho}_x$ as the coupling form, which could be successfully free from the large scale instability of the perturbations. In section III, we show the effects on the CMB temperature power spectra, matter power spectra, and evolution curves of $f\sigma_8(z)$ for varied interaction rate. Then, we conduct a global fitting on the coupled dark energy model by the Markov Chain Monte Carlo (MCMC) approach, and discuss the results of the cosmological constraints, particularly, with respect to the interaction rate. In the final section, our conclusions are presented.

\section{The perturbation equations of dark fluids}

In a flat universe described by the FRW metric, we parameterize the interaction as follows
\begin{eqnarray}
\nabla_{\nu}T^{\mu\nu}_{(x)}= Qu^{\mu}_{(x)}, \\
\nabla_{\nu}T^{\mu\nu}_{(c)}=-Qu^{\mu}_{(x)},
\label{eq:balance-equation}
\end{eqnarray}
where $T^{\mu\nu}_{(c)}=\bar{T}^{\mu\nu}_{(c)}+\delta T^{\mu\nu}_{(c)}$ and $T^{\mu\nu}_{(x)}=\bar{T}^{\mu\nu}_{(x)}+\delta T^{\mu\nu}_{(x)}$ are the energy-momentum tensors for the dark matter and dark energy, respectively. $u^{\mu}_{(x)}$ is the four-velocity of dark energy, which means that the energy transfer four-vector is parallel to the four-velocity of dark energy \cite{ref:Clemson2012,ref:Koyama2009}. $Q$ is the scalar energy transfer rate between the dark components. Here, we consider that the energy transfer rate is proportional to the Hubble parameter and the energy density of dark energy, that is, $Q=\bar{Q}+\delta Q=3\xi_xH\bar{\rho}_x(1+\delta_x)$. This coupling form is free of large scale instability \cite{ref:Valiviita2008,ref:Clemson2012,ref:Gavela2009}. Furthermore, in the light of Refs. \cite{ref:Gavela2009,ref:Clemson2012}, the stability conditions of the perturbations are $\xi_x>0$ and $(1+w_x)>0$. As for the phantom case $w_x<-1$, together with $\xi_x<0$, which does not suffer from the instability, but we exclude it on account of the unphysical property \cite{ref:Caldwell2003}. For the momentum transfer potential, we assume it as zero in the rest frame of either dark matter or dark energy \cite{ref:Valiviita2008,ref:Koyama2009}. This leads to two cases of simple interacting model which include energy transfer four-vector parallel to the four-velocity of dark matter or dark energy \cite{ref:Clemson2012,ref:Koyama2009}.

In the synchronous gauge, for the $\xi$wCDM model, the continuity and Euler equations of the dark energy and dark matter read \cite{ref:Yang2014-uc}
\begin{eqnarray}
\delta'_x+(1+w_x)\left(\theta_x+\frac{h'}{2}\right)+3\mathcal{H}(c^2_{sx}-w_x)\delta_x
+9\mathcal{H}^2(1+w_x)(c^2_{sx}-w_x)\frac{\theta_x}{k^2}
=9\mathcal{H}^2(c^2_{sx}-w_x)\xi_x\frac{\theta_x}{k^2},
\label{eq:deltax-prime-b}
\end{eqnarray}
\begin{eqnarray}
\delta'_c+\theta_c+\frac{h'}{2}=3\mathcal{H}\xi_x\frac{\bar{\rho}_x}{\bar{\rho}_c}(\delta_c-\delta_x),
\label{eq:deltac-prime-b}
\end{eqnarray}
\begin{eqnarray}
\theta'_x+\mathcal{H}(1-3c^2_{sx})\theta_x-\frac{c^2_{sx}}{1+w_x}k^2\delta_x
=\frac{3\mathcal{H}\xi_x}{1+w_x}[b(\theta_c-\theta_x)-c^2_{sx}\theta_x],
\label{eq:thetax-prime-b}
\end{eqnarray}
\begin{eqnarray}
\theta'_c+\mathcal{H}\theta_c
=3\mathcal{H}\xi_x\frac{\bar{\rho}_x}{\bar{\rho}_c}(1-b)(\theta_c-\theta_x),
\label{eq:thetac-prime-b}
\end{eqnarray}
where $c^2_{sx}$ is the physical sound speed of dark energy in the rest frame, its definition is $c^2_{sx}=\delta p_x/\delta\rho_x|_{restframe}$ \cite{ref:Valiviita2008,ref:Kodama1984,ref:Hu1998,ref:Gordon2004}. In order to avoid the unphysical instability, $c^2_{sx}$ should be taken as a non-negative parameter \cite{ref:Valiviita2008}. In the above equations, $b=1$ corresponds to the coupled model with $Q^{\mu}\parallel u^{\mu}_{(c)}$ \cite{ref:Yang2014-uc}; $b=0$ corresponds to the coupled model with $Q^{\mu}\parallel u^{\mu}_{(x)}$ in this paper.

\section{Cosmological implications and constraint results}

\subsection{Theoretical predictions of CMB temperature, matter power spectra, and $f\sigma_8(z)$ evolution}

\begin{figure}[!htbp]
\includegraphics[width=8.8cm,height=6.8cm]{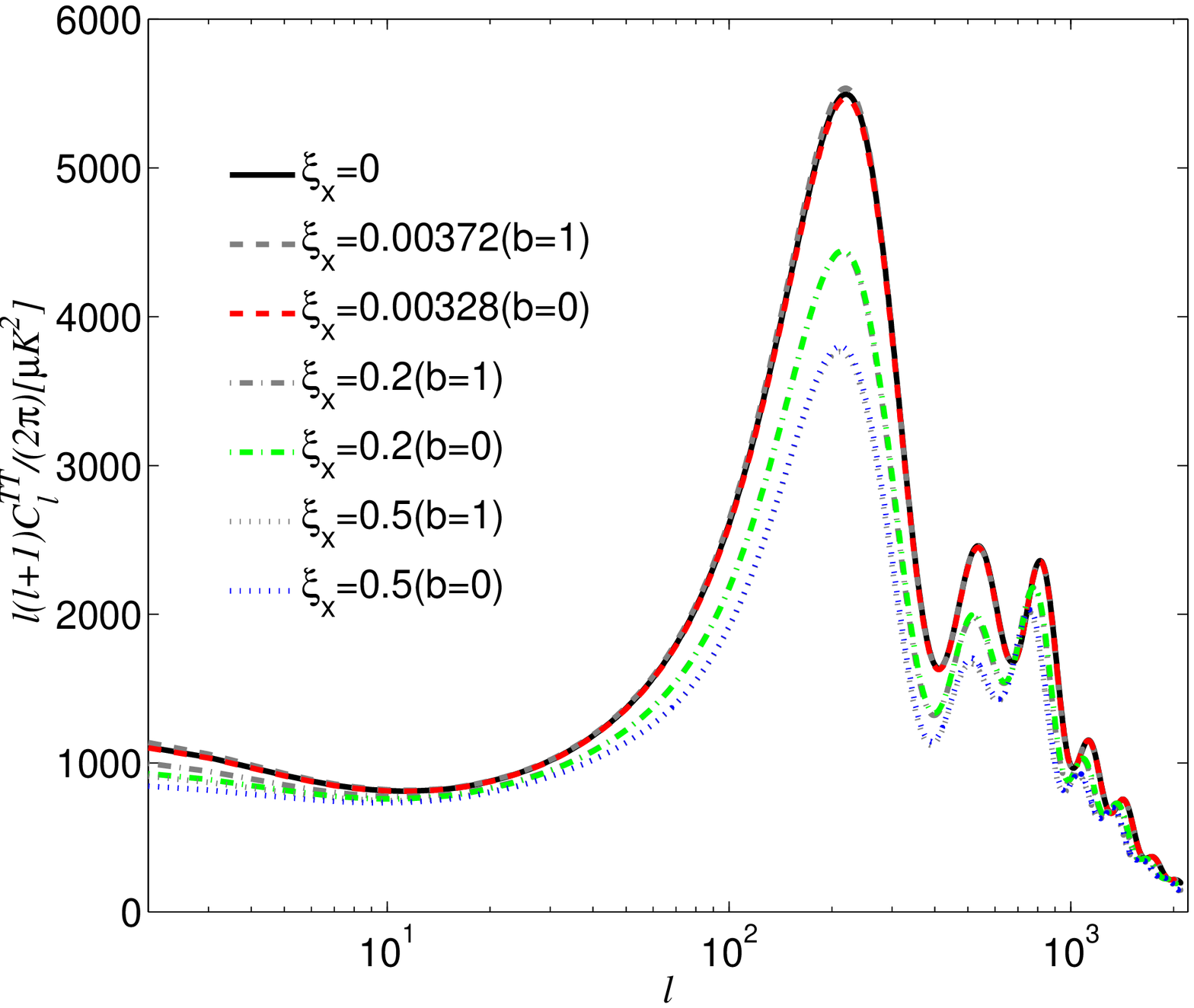}
\includegraphics[width=8.8cm,height=6.8cm]{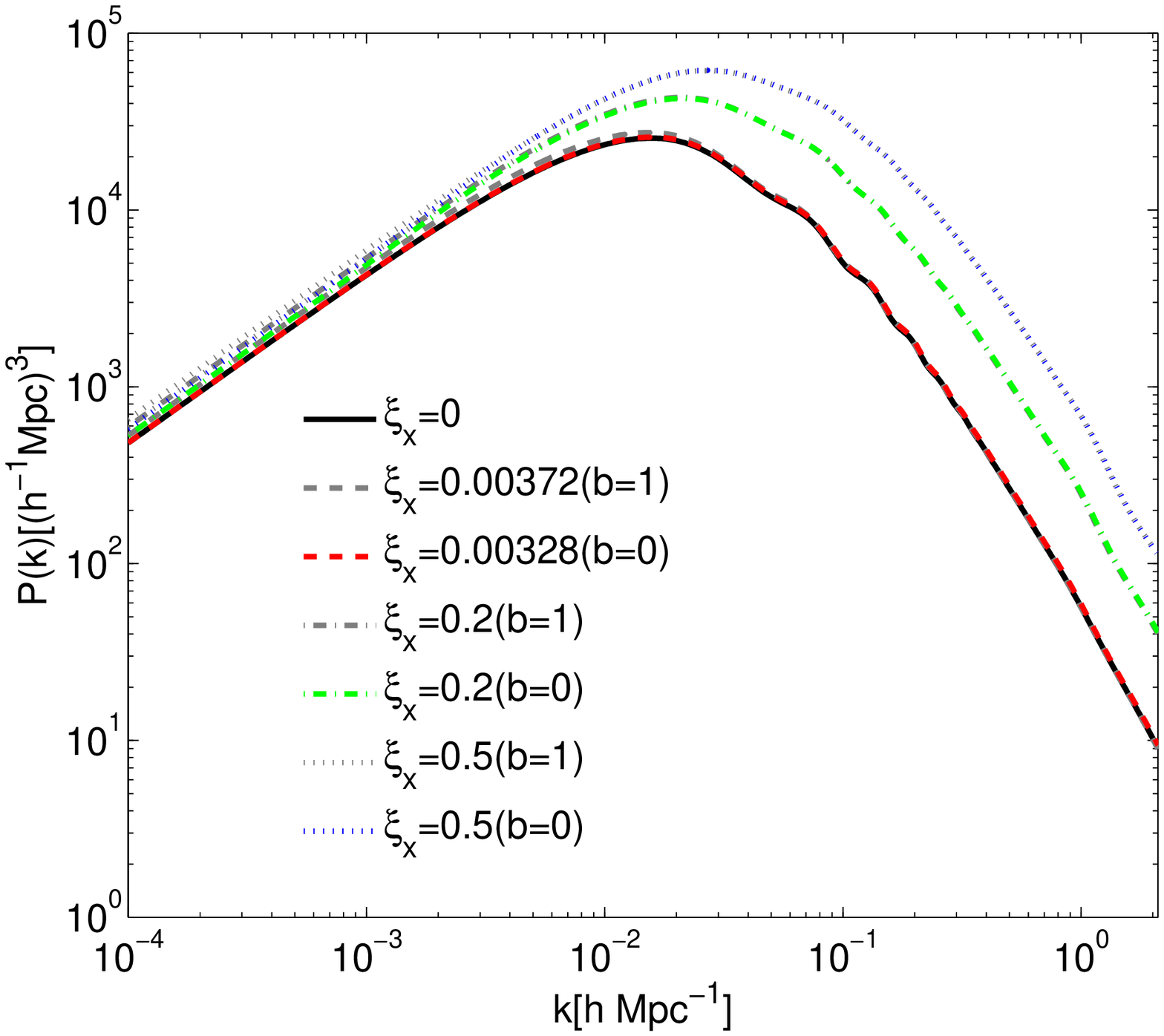}
  \caption{The effects on the CMB temperature power spectra (left panel) and matter power spectra (right panel) for the different values of interaction rate $\xi_x$. The black solid, red thick dashed, green dotted-dashed, and blue dotted lines are for $\xi_x=0, 0.00328, 0.2$, and $0.5$, respectively; the other relevant parameters are fixed with the mean values as shown in the fourth column of Table \ref{tab:results-mean-uxpos}. The gray dashed, dotted-dashed, and dotted lines are the corresponding ones for the coupled model with $Q^{\mu}\parallel u^{\mu}_{(c)}$ ($b=1$).}
  \label{fig:CMBpower-Mpower}
\end{figure}

When the coupling case $Q^{\mu}\parallel u^{\mu}_{(x)}$ ($b=0$) happened, in the modified \textbf{CAMB} code, we fix some input parameters and only vary the interaction rate $\xi_x$ so that we could gain an insight into the physical implications of the coupling between the dark components. The two panels of Fig. \ref{fig:CMBpower-Mpower} show the CMB temperature and matter power spectra for four values of $\xi_x$ with all other cosmological parameters set to mean values in the fourth column of Table \ref{tab:results-mean-uxpos}. Positive $\xi_x$ presents that the direction of energy transfer is from dark matter to dark energy, when the fractional dimensionless density parameter $\Omega_c$ is fixed today, the energy density of dark matter would have been correspondingly greater in the past than without the uncoupled model. It means that a larger proportion of dark matter and a more significant effect from photon driving before decoupling, which could make the amplitude of the CMB temperature power spectra decreased and the position of the peaks shifted. Moreover, the deviation from the standard evolution law $a^{-3}$ of dark matter affects the CMB temperature power spectra at the low $l$ part via the integrated Sachs-Wolfe (ISW) effect due to the evolution of gravitational potential. The present matter power spectra represent a relative increase for varied interaction rate $\xi_x$, since in the past the more energy density of dark matter naturally leads to more structure formation and an increase in the amplitude of the matter power spectra.
We also show the CMB temperature and matter power spectra of the coupled model with $Q^{\mu}\parallel u^{\mu}_{(c)}$ ($b=1$) in Fig. \ref{fig:CMBpower-Mpower}, and do not find significant differences between the two coupling cases. The CMB temperature power spectra at the low-$l$ part and matter power spectra at very large scale have slight differences.

\subsection{Modified growth of structure}

We consider dark energy does not cluster on sub-Hubble scales \cite{ref:Koyama2009,ref:Clemson2012}, we could ignore the term about $\delta_x$ in Eq. (\ref{eq:deltac-prime-b}) and obtain the second-order differential equation of density perturbation about dark matter

\begin{eqnarray}
\delta''_c+\left(1-6\xi_x\frac{\bar{\rho}_x}{\bar{\rho}_c}\right)\mathcal{H}\delta'_c
=4\pi Ga^2\bar{\rho}_b\delta_b + 4\pi Ga^2\bar{\rho}_c\delta_c \left[1+ 2\xi_x\frac{\bar{\rho}_{tot}}{\bar{\rho}_c}\frac{\bar{\rho}_x}{\bar{\rho}_c}
\left( \frac{\mathcal{H}'}{\mathcal{H}^2}+1-3w_x+3\xi_x \right) \right]
+3\xi_x\frac{\bar{\rho}_x}{\bar{\rho}_c}\theta_x,
\label{eq:thetac-prime2-ide2-mod}
\end{eqnarray}
where $\mathcal{H}^2=8\pi Ga^2\bar{\rho}_{tot}/3$. During matter domination, the velocity perturbations of dark energy $\theta_x$ is small enough to be negligible. When $\xi_x$=0, the above equation could be turned into the standard evolution of matter perturbations $\ddot{\delta}_m+2H\dot{\delta}_m-4\pi G(\delta\rho+3\delta p)=0$ \cite{ref:Linder2003}.

\begin{figure}[!htbp]
\includegraphics[width=8.5cm,height=6.6cm]{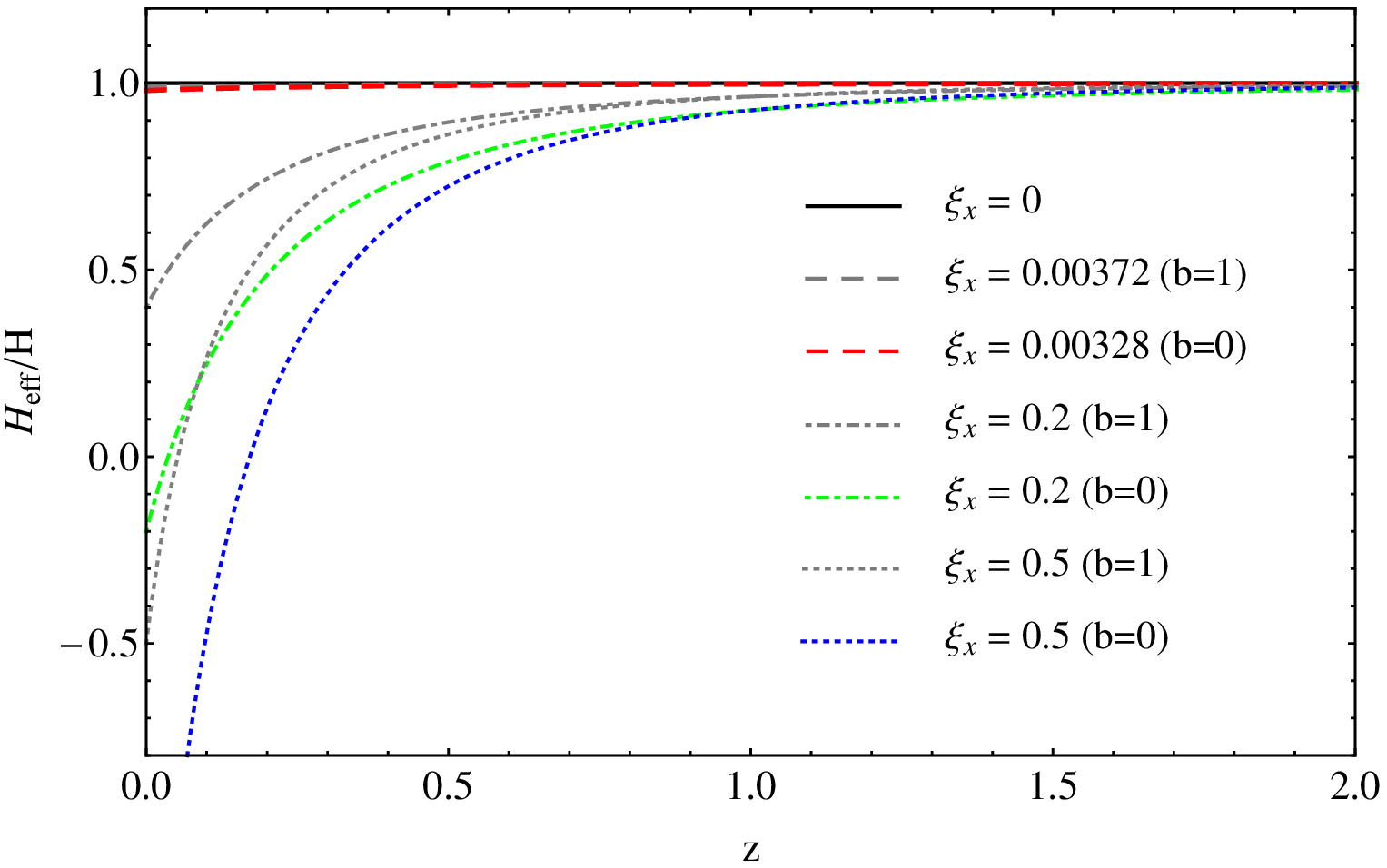}
\includegraphics[width=8.5cm,height=6.6cm]{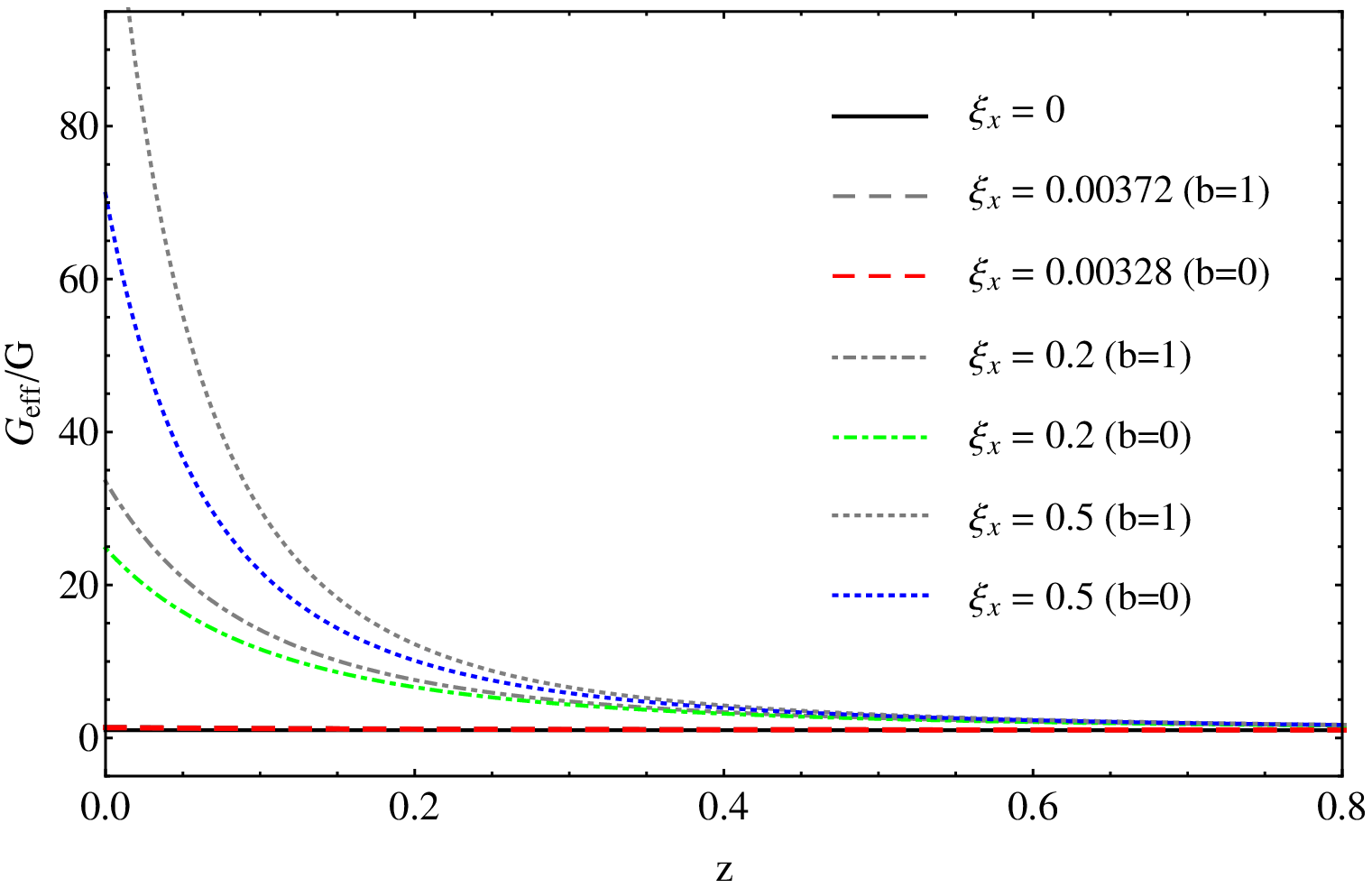}
  \caption{Deviations from standard model of the effective Hubble parameter (left panel) and effective gravitational constant (right panel) for $\delta_c$. The black solid, red thick dashed, green dotted-dashed, and blue dotted lines are for $\xi_x=0, 0.00328, 0.2$, and $0.5$, respectively; $\xi_x=0$ corresponds to the case of $\Lambda$CDM model; the other relevant parameters are fixed with the mean values as shown in the fourth column of Table \ref{tab:results-mean-uxpos}. The gray dashed, dotted-dashed, and dotted lines are the corresponding ones for the coupled model with $Q^{\mu}\parallel u^{\mu}_{(c)}$ ($b=1$), where $\frac{\mathcal{H}_{eff}}{\mathcal{H}}|_{(b=1)}=1-3\xi_x\frac{\bar{\rho}_x}{\bar{\rho}_c}$ and $\frac{G_{eff}}{G}|_{(b=1)}=1+ 2\xi_x\frac{\bar{\rho}_{tot}}{\bar{\rho}_c}\frac{\bar{\rho}_x}{\bar{\rho}_c}
\left[ \frac{\mathcal{H}'}{\mathcal{H}^2}+1-3w_x+3\xi_x\left(1+\frac{\bar{\rho}_x}{\bar{\rho}_c}\right) \right]$.}
  \label{fig:Heff-Geff}
\end{figure}

The evolutions of $\delta_c$ for the coupled model also bring about the deviations from the standard model from another two aspect. The first one is the modified Hubble friction term $\mathcal{H}_{eff}$; The second one is the modified source term, which is from the modified effective gravitational constant $G_{eff}$.
We present the evolutions of $\mathcal{H}_{eff}/\mathcal{H}$ and $G_{eff}/G$ in Fig. \ref{fig:Heff-Geff}. When the interaction rate is large, from the early times to the late times, the evolution of $\mathcal{H}_{eff}/\mathcal{H}$ shows exponential decreasement, meanwhile, the evolution of $G_{eff}/G$ takes on exponential increasement.
Furthermore, we find the obvious differences from $\mathcal{H}_{eff}/\mathcal{H}$ and $G_{eff}/G$ between the two coupled models with $b=0$ and $b=1$, particularly, when the interaction rate obviously deviates from zero.

\begin{eqnarray}
\frac{\mathcal{H}_{eff}}{\mathcal{H}}=1-6\xi_x\frac{\bar{\rho}_x}{\bar{\rho}_c},
\label{eq:Heff}
\end{eqnarray}

\begin{eqnarray}
\frac{G_{eff}}{G}=1+ 2\xi_x\frac{\bar{\rho}_{tot}}{\bar{\rho}_c}\frac{\bar{\rho}_x}{\bar{\rho}_c}
\left( \frac{\mathcal{H}'}{\mathcal{H}^2}+1-3w_x+3\xi_x \right).
\label{eq:Geff}
\end{eqnarray}

%Here, there is an important issue about the growth rate for the coupled case $Q^{\mu}\parallel u^{\mu}_{(x)}$.
As usual, we define the growth rate as $f_m=d\ln\delta_m/d\ln a$. In fact, this is because that the Euler equation is unchanged,
% such as the coupled case $Q^{\mu}\parallel u^{\mu}_{(c)}$ \cite{ref:Yang2014-uc},
the dark matter could follow geodesics. If the Euler equation is modified,
% such as the coupled case $Q^{\mu}\parallel u^{\mu}_{(x)}$,
the dark matter no longer follows geodesics in general \cite{ref:Koyama2009}. In this case, we choose a gauge invariant variable $\Delta_m=\delta_m+3\mathcal{H}\theta_m/k^2$, and define the growth rate according to $f_m=d\ln\Delta_m/d\ln a$ \cite{ref:Bean2010}. Moreover, we should pay attention to $\Delta_c$, which is related to both the scale factor $a$ and the wavenumber $k$. This alerts us to notice the scale-dependence of the RSD measurement, though Refs. \cite{ref:fsigma8total-Samushia2013,ref:fsigma84-Reid2012,ref:Yang2014-uc} have claimed that the linear growth is scale-independent in the theoretical frame of general relativity. From the relation $\Delta_m=(\bar{\rho}_c/\Delta_c+\bar{\rho}_b\Delta_b)/(\bar{\rho}_c+\bar{\rho}_b)$, the matter density perturbations could be written as
\begin{eqnarray}
\Delta_m=\frac{\bar{\rho}_c\delta_c+\bar{\rho}_b\delta_b}{\bar{\rho}_c+\bar{\rho}_b}
+\frac{3}{k^2}\mathcal{H}\theta_c\frac{\bar{\rho}_c}{\bar{\rho}_c+\bar{\rho}_b}.
\label{eq:deltam-capital}
\end{eqnarray}

In order to clearly show the difference between the two coupling cases $Q^{\mu}\parallel u^{\mu}_{(x)}$ and $Q^{\mu}\parallel u^{\mu}_{(c)}$, we also present the corresponding matter density perturbations for $b=1$,
\begin{eqnarray}
\delta_m|_{(b=1)}=\frac{\bar{\rho}_c\delta_c+\bar{\rho}_b\delta_b}{\bar{\rho}_c+\bar{\rho}_b}.
\end{eqnarray}

Moreover, the growth rate reads
\begin{eqnarray}
f_m=\frac{d\ln\Delta_m}{d\ln a}=\frac{1}{\mathcal{H}}\frac{\Delta'_m}{\Delta_m},
\label{eq:fm-capital}
\end{eqnarray}
where
\begin{eqnarray}
\Delta'_m=\frac{3\mathcal{H}[(\xi_x\bar{\rho}_x+\bar{\rho}_c+\bar{\rho}_b)\delta_m
-(\xi_x\bar{\rho}_x\delta_c+\bar{\rho}_c\delta_c+\bar{\rho}_b\delta_b)]
+\bar{\rho}_c\delta'_c+\bar{\rho}_b\delta'_b}{\bar{\rho}_c+\bar{\rho}_b}
+\frac{3}{k^2}\theta_c\frac{\bar{\rho}_c}{\bar{\rho}_c+\bar{\rho}_b}
\left[ \mathcal{H}'-\mathcal{H}^2
-\frac{3\mathcal{H}^2\xi_x\bar{\rho}_x\bar{\rho}_b}{(\bar{\rho}_c+\bar{\rho}_b)\bar{\rho}_c} \right]
\nonumber.
\label{eq:deltam-capital-prime}
\end{eqnarray}

The growth rate is different from the one of $Q^{\mu}\parallel u^{\mu}_{(c)}$,
\begin{eqnarray}
f_m|_{(b=1)}=\frac{d\ln\delta_m}{d\ln a}=\frac{1}{\mathcal{H}}\frac{\delta'_m}{\delta_m},
\end{eqnarray}
\begin{eqnarray}
\delta'_m|_{(b=1)}=\frac{3\mathcal{H}[(\xi_x\bar{\rho}_x+\bar{\rho}_c+\bar{\rho}_b)\delta_m
-(\xi_x\bar{\rho}_x\delta_c+\bar{\rho}_c\delta_c+\bar{\rho}_b\delta_b)]
+\bar{\rho}_c\delta'_c+\bar{\rho}_b\delta'_b}{\bar{\rho}_c+\bar{\rho}_b}. \nonumber
\end{eqnarray}

\begin{figure}[!htbp]
\includegraphics[width=8.8cm,height=6.8cm]{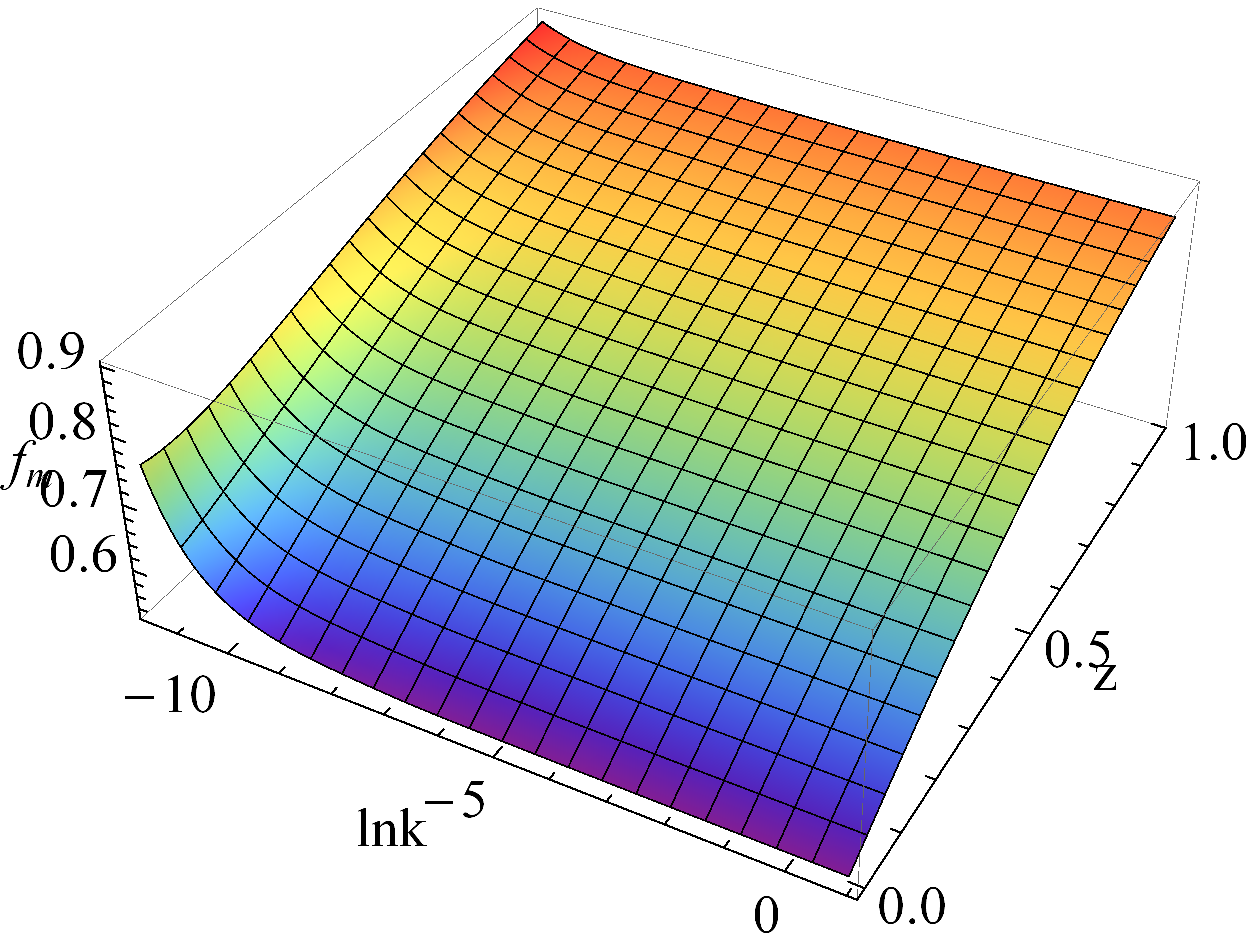}
\includegraphics[width=8.8cm,height=6.8cm]{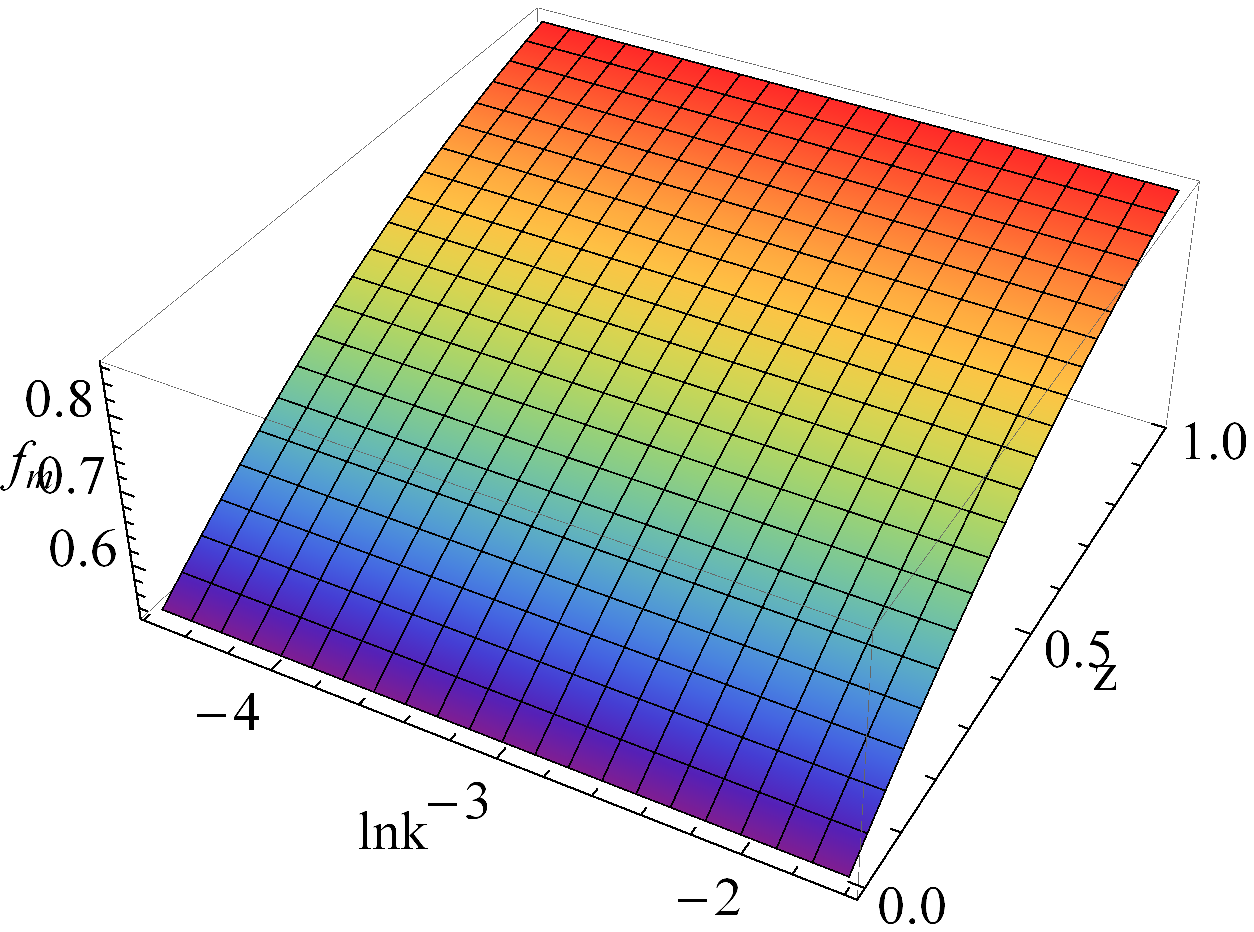}
%  \caption{The three-dimensional plots of $\ln k$ ($k$ is the wavenumber), $z$ (redshift), $f_m$ (growth rate of matter). The left panel corresponds to the coupled case of $Q^{\mu}\parallel u^{\mu}_{(x)}$; the right panel corresponds to the coupled case of $Q^{\mu}\parallel u^{\mu}_{(c)}$, which is from Fig. 6 of Ref. \cite{ref:Yang2014-uc}. In the left panel, $\xi_x$ and the other relevant parameters are fixed with the mean values as shown in the fourth column of Table \ref{tab:results-mean-uxpos}.}
  \caption{The three-dimensional plots of $\ln k$ ($k$ is the wavenumber), $z$ (redshift), $f_m$ (growth rate of matter) for different ranges of the wavenumber. In the two panels, $\xi_x$ and the other relevant parameters are fixed with the mean values as shown in the fourth column of Table \ref{tab:results-mean-uxpos}.}
  \label{fig:growth-kzf}
\end{figure}

From the expressions of $\Delta_m$ and $\Delta'_m$, we know the growth rate is scale-dependent, which is different from the coupled model with $Q^{\mu}\parallel u^{\mu}_{(c)}$ in Ref. \cite{ref:Yang2014-uc}. Particularly, from the second terms of $\Delta_m$ and $\Delta'_m$, it is easy to see that the growth history would be significantly affected, when the wavenumber $k$ is small enough. According to $f_m=\Delta'_m/(\mathcal{H}\Delta_m)$, one could modify the \textbf{CAMB} code \cite{ref:camb}, and put the wavenumber $k$, redshift $z$, and growth rate $f_m$ into a three-dimensional table.
To clearly show the modified growth history, we present the three-dimensional plots of $k$, $z$, $f_m$ in the left panel of Fig. \ref{fig:growth-kzf}.
Obviously, the growth rate is scale-dependent when $k$ is small enough. Fortunately, the analysis of redshift-space distortion is valid in the range of the wavenumber $k\in(0.01,0.20)[hMpc^{-1}]$. In this range, we also show the three-dimensional plots in the right panel of Fig. \ref{fig:growth-kzf}, the growth rate is almost scale-independent. Therefore, we could safely obtain the model parameter space with the $f\sigma_8(z)$ data points.
In order to strongly confirming this opinion from the cosmological constraint, we would constrain the coupled model for several different scales in the next section.

\begin{figure}[!htbp]
%$k\in(0.6,0.8)[hMpc^{-1}]$
\includegraphics[width=13cm,height=9cm]{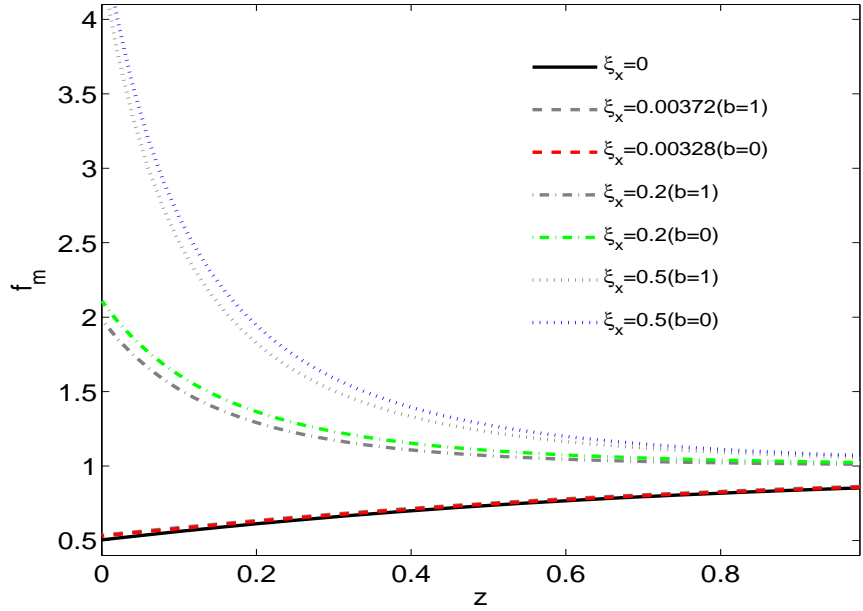}
  \caption{The evolutions for the growth rate of the matter for $k=0.065[hMpc^{-1}]$. The black solid, red thick dashed, green dotted-dashed, and blue dotted lines are for $\xi_x=0, 0.00328, 0.2$, and $0.5$, respectively; $\xi_x=0$ corresponds to the case of standard model; the other relevant parameters are fixed with the mean values as shown in the fourth column of Table \ref{tab:results-mean-uxpos}. The gray dashed, dotted-dashed, and dotted lines are the corresponding ones for the coupled model with $Q^{\mu}\parallel u^{\mu}_{(c)}$ ($b=1$).}
  \label{fig:fz}
\end{figure}

Modified evolution of the matter perturbation $\Delta_m$ determines that the growth history would deviate from the standard case in the theoretical frame of general relativity. The evolutions of growth rate are shown in Fig. \ref{fig:fz}. From this figure, we clearly see that the interaction rate $\xi_x$ could significantly affect the growth history of the Universe, the growth rate presents large differences at late times. It means that the growth history of dark matter is significantly sensitive to the varied interaction rate.
Moreover, in Fig. \ref{fig:fz}, we find the obvious differences from the growth rate between the two coupling cases, particularly, when the interaction rate obviously deviates from zero. Why are they different? We should look for the answer from the Euler equation (\ref{eq:thetac-prime-b}), we follow the discussion of Refs. \cite{ref:Koyama2009,ref:Clemson2012} and analyze the velocity bias for the coupled model with $b=1$.

Here, we recall the Euler equation of the baryons
%\begin{eqnarray}
%\theta'_c+\mathcal{H}\theta_c-k^2\phi
%=-\frac{aQ_c}{\rho_c}(1-b)(\theta_c-\theta_x),
%\label{eq:general-thetac-b2}
%\end{eqnarray}
\begin{eqnarray}
\theta'_b+\mathcal{H}\theta_b=0.
\label{eq:thetac-prime-b3}
\end{eqnarray}

From the above equation and Euler equation of dark matter (\ref{eq:thetac-prime-b}), it is easy to see that the Euler equation of dark matter for the coupling case $b=1$ is the same as the uncoupled model, that is, there is no dark matter-baryon velocity difference. However, for the coupled model with $b=0$, the right-hand side of the Euler equation (\ref{eq:thetac-prime-b}) is nonzero, then dark matter no longer follows the geodesics and breaks the weak equivalence principle \cite{ref:Koyama2009}. In the coupled dark energy model with $b=0$, the Euler equations of dark matter and baryons imply
\begin{eqnarray}
(\theta_c-\theta_b)'+\left( 1-3\xi_x\frac{\bar{\rho}_x}{\bar{\rho}_c} \right)\mathcal{H}(\theta_c-\theta_b)
=-3\mathcal{H}\xi_x\frac{\bar{\rho}_x}{\bar{\rho}_c}(\theta_x-\theta_b).
\label{eq:thetac-prime-b4}
\end{eqnarray}

Therefore, there would be the velocity bias between dark matter and baryons in the coupled model with $Q^{\mu}\parallel u^{\mu}_{(x)}$. Modified Euler equation brings about the different structure formation histories for the two coupled models. In the coupled model with $b=0$, the dark matter receives a change in the momentum from the dark energy perturbations, as expressed by its modified Euler equation (\ref{eq:thetac-prime-b}), leading to more structure growth at late times relative to the coupled model with $b=1$ (we could see it from Fig. \ref{fig:fz}). At the same time, dark energy would be weaker for the coupled model with $b=0$ at late times, different growth rates lead to correspondingly dissimilar ISW signatures. Therefore, when the interaction rate of the two coupled models is fixed on the same value, such as 0.2 or 0.5, at the low$-l$ region of present CMB temperature power spectra in Fig. \ref{fig:CMBpower-Mpower}, the ISW signal is slightly lower relative to the coupled model with $b=1$.

Since the growth history of the coupled dark energy model is obviously sensitive to the interaction rate, it is worthwhile to anticipate that the large scale structure measurement will help to significantly tighten the cosmological constraints and some evidences could rule out large interaction rate. In order to adopt the RSD measurement, we add a new module \textbf{CosmoMC} package \cite{ref:cosmomc-Lewis2002} to import $f_m$ from CAMB which could be used to calculate the theoretical values of $f\sigma_8(z)$ at ten different redshifts. For constraining the other cosmological models with the RSD analysis, please see Refs. \cite{ref:Yang2014-uc,ref:fsigma8-HDE-Xu2013,ref:fsigma8andPlanck-MG-Xu2013,ref:Xu2013-DGP,ref:Yang2013-wdm,ref:Yang2013-cass}.

\begin{figure}[!htbp]
\includegraphics[width=13cm,height=9cm]{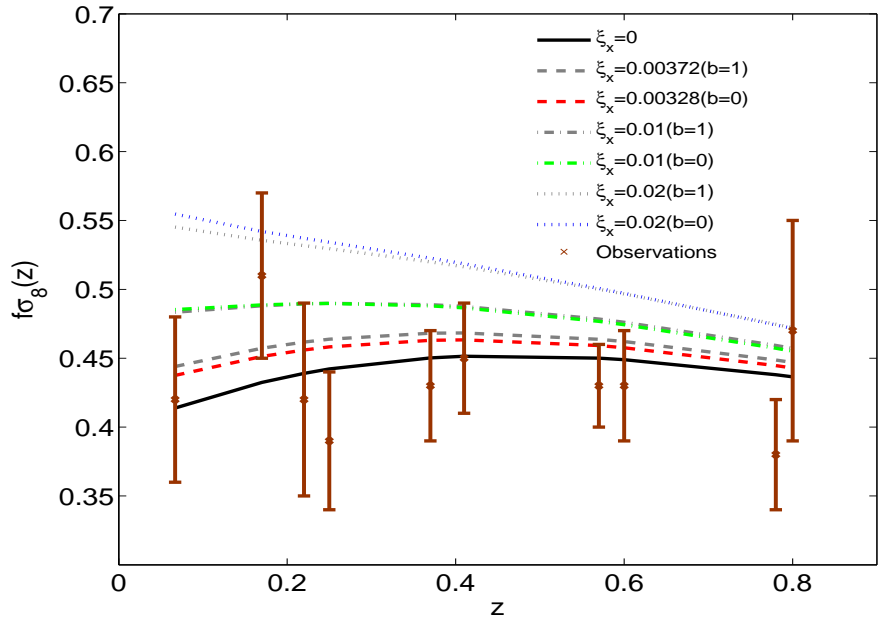}
  \caption{The fitting evolutionary curves of $f\sigma_8(z)$ about the redshift $z$ for varied interaction rate $\xi_x$. The black solid, red thick dashed, green dotted-dashed, and blue dotted lines are for $\xi_x=0, 0.00328, 0.01$, and $0.02$, respectively; the brown error bars denote the $f\sigma_8(z)$ observations at different redshifts, which are listed in Table \ref{tab:fsigma8data}; the other relevant parameters are fixed with the mean values as shown in the fourth column of Table \ref{tab:results-mean-uxpos}. The gray dashed, dotted-dashed, and dotted lines are the corresponding ones for the coupled model with $Q^{\mu}\parallel u^{\mu}_{(c)}$ ($b=1$).}
  \label{fig:fsigma8-uxpos}
\end{figure}

Furthermore, in Fig. \ref{fig:fsigma8-uxpos}, we also plot the evolutions of $f\sigma_8(z)$ when the interaction rate is varied. Positive interaction rate denotes a transfer of energy from dark matter to dark energy, with fixed $\Omega_c$ today, the dark matter energy density would be greater in the past than the uncoupled model. A larger proportion of dark matter naturally leads to more structure growth (as is shown in Fig. \ref{fig:fz}) and the increase of present matter power spectra (as is shown in the right panel of Fig. \ref{fig:CMBpower-Mpower}), which are correspondingly the larger growth rate and the higher $\sigma_8$ ($\sigma_8$ could be obtained by the integration with regard to the matter power spectra \cite{ref:fsigma8-DE-Song2009,ref:Percival2009}). Therefore, the values of $f\sigma_8(z)$ are enhanced than the uncoupled model, and the amplitude of enhancement becomes obvious with raising the values of $\xi_x$. Besides, from Eqs. (\ref{eq:thetac-prime2-ide2-mod}), (\ref{eq:Heff}), and (\ref{eq:Geff}), we also could know why the changed amplitude of $f\sigma_8(z)$ becomes large with reducing the redshift. For fixed $\xi_x$, at the higher redshift, the component of dark energy is subdominant, the modified Hubble friction term and source term are trivial, which would slightly affect the evolutions of growth rate and $\sigma_8$. Nonetheless, at the lower redshift, the dark energy gradually dominate the late Universe, the modified $H_{eff}$ and $G_{eff}$ would significantly increase the cosmic structure growth, which could bring about more obvious enhancement on the evolutions of $f\sigma_8(z)$.
Besides, the evolutions of $f\sigma_8(z)$ are slightly different from the coupling case $Q^{\mu}\parallel u^{\mu}_{(c)}$ ($b=1$). The reason is that the two coupled models undergo different growth histories, dissimilar growth rates lead to different evolutions of $f\sigma_8(z)$. At late times, that is, at the lower redshift, with increasing the interaction rate, the difference of the $f\sigma_8(z)$ evolutions between the two models become relatively obvious.
Furthermore, it is easy to see that the case of $\xi_x=0.00328$ (corresponds the IwCDM model with mean value) and that of $\xi_x=0$ (corresponds to the uncoupled wCDM model) are differentiable from the evolution curves of $f\sigma_8(z)$, which is different from the evolutions of CMB temperature and matter power spectra. It means that, to some extent, the RSD test could break the possible degeneracy between the IwCDM model and the uncoupled wCDM model.
Besides, in Figs. \ref{fig:fz} and \ref{fig:fsigma8-uxpos}, from the comparison between the two coupled models with $b=0$ and $b=1$, it turns out that there actually are differences on the modified growth history between the two models, however, it is also clear that it will be vary hard to distinguish them.

\subsection{Data sets and constraint results}

To constrain the interaction rate from the currently available cosmic observations, we adopt the MCMC method to perform the likelihood analysis under the stability conditions $\xi_x>0$ and $(1+w_x)>0$. We do this by making use of the publicly available \textbf{CosmoMC} package. The code is modified in the light of the perturbation evolutions of the $\xi$wCDM model for the case of $b=0$. We consider the following eight-dimensional parameter space for the coupled model
\begin{eqnarray}
P\equiv\{\Omega_bh^2, \Omega_{c}h^2, \Theta_S, \tau, w_x, \xi_x, n_s, log[10^{10}A_S]\},
\label{eq:parameter_space}
\end{eqnarray}
where $\Omega_bh^2$ and $\Omega_{c}h^2$, respectively, stand for the density of the baryons and dark matter, $\Theta_S=100\theta_{MC}$ refers to the ratio of sound horizon and angular diameter distance, $\tau$ indicates the optical depth, $w_x$ is the EoS of dark energy, $\xi_x$ is the interaction rate between the dark components, $n_s$ is the scalar spectral index, and $A_s$ represents the amplitude of the initial power spectrum. The priors to the basic model parameters are listed in the second column of Table \ref{tab:results-mean-uxpos-3k}. Here, the pivot scale of the initial scalar power spectrum $k_{s0}=0.05Mpc^{-1}$ is used. Then, based on the MCMC method, we perform a global fitting for the coupled model with $Q^{\mu}\parallel u^{\mu}_{(x)}$ when the model parameters satisfy $\xi_x>0$ and $(1+w_x)>0$. Here, we choose $c^2_{sx}=1$ which could avoid the unphysical sound speed \cite{ref:Valiviita2008,ref:Majerotto2010,ref:Clemson2012}.

For our numerical calculations, the total likelihood $\chi^2$ can be constructed as
\begin{eqnarray}
\chi^2=\chi^2_{CMB}+\chi^2_{BAO}+\chi^2_{SNIa}+\chi^2_{RSD},
\label{eq:chi2}
\end{eqnarray}
where the used data sets for our MCMC likelihood analysis are listed in Table \ref{tab:alldata}. Some detailed descriptions about the observed data sets have been shown in Appendix C of Ref. \cite{ref:Yang2014-uc}.

\begin{table}
\begin{center}
\begin{tabular}{|c|c|}
\hline\hline Data names & Data descriptions and references \\ \hline
CMB & $l\in[50,2500]$ temperature likelihood from Planck \cite{ref:Planck2013-params} \\
$...$ & up to $l=49$ temperature likelihood from Planck \cite{ref:Planck2013-params} \\
$...$ & up to $l=32$ polarization likelihood from WMAP9 \cite{ref:WMAP9} \\
BAO & $r_s/D_V(z=0.106)=0.336\pm0.015$ \cite{ref:BAO-1}\\
$...$ & $r_s/D_V(z=0.35)=0.1126\pm0.0022$ \cite{ref:BAO-2}\\
$...$ & $r_s/D_V(z=0.57)=0.0732\pm0.0012$ \cite{ref:BAO-3}\\
SNIa & SNLS3 data from SiFTO and SALT2 \cite{ref:SNLS3-1,ref:SNLS3-2,ref:SNLS3-3}\\
RSD & ten $f\sigma_8(z)$ data points from Table \ref{tab:fsigma8data}\\
\hline\hline
\end{tabular}
\caption{The used data sets for our MCMC likelihood analysis on the coupled dark energy model, where $l$ is the multipole number of power spectra, and WMAP9 is the abbreviation of nine-year Wilkinson Microwave Anisotropy Probe.}
\label{tab:alldata}
\end{center}
\end{table}

\begin{figure}[!htbp]
\includegraphics[width=14cm,height=10cm]{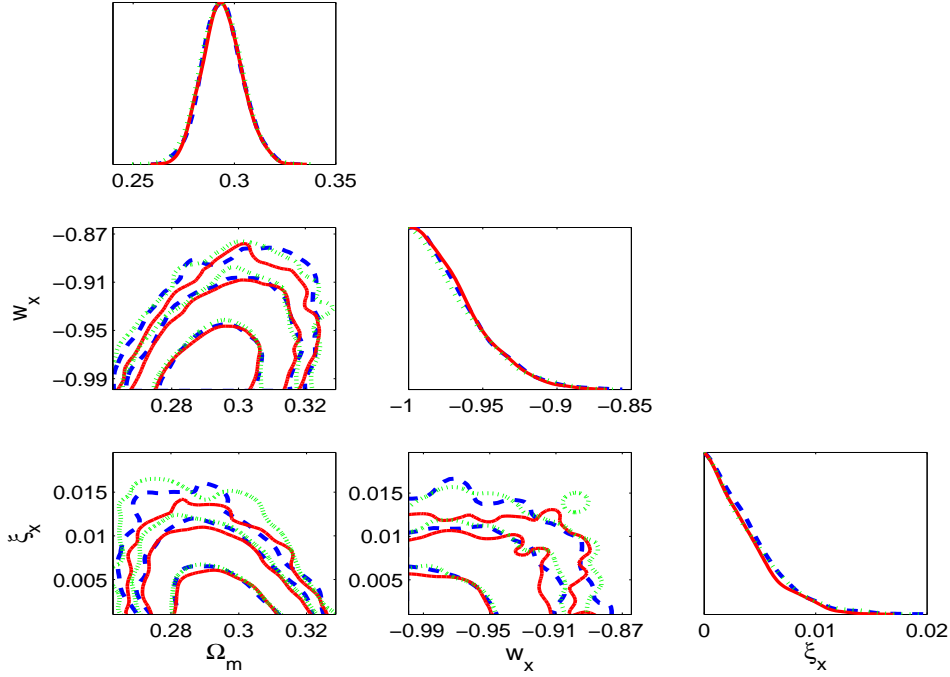}
  \caption{For several different scales, the 1D marginalized distributions on the parameters ($\Omega_m$, $w_x$, $\xi_x$) and 2D contours between each other with 68\% C.L., 95 \% C.L., and 99.7\% C.L., where CMB from Planck + WMAP9, BAO, SNIa, and RSD data sets have been used. The green dotted, red solid, and blue dashed lines are for $k=0.015$, $0.065$, and $0.165[hMpc^{-1}]$, respectively.}
  \label{fig:contour-uxpos-3k}
\end{figure}

For several different scales, we have constraint the coupled model with CMB, BAO, SN, and RSD. The mean values with $1,2\sigma$ errors and best-fit values of the parameters were presented in Table \ref{tab:results-mean-uxpos-3k} after running eight chains in parallel on the computer. In Figs. \ref{fig:contour-uxpos-3k}, we also show the one-dimensional (1D) marginalized distributions of the parameters ($\Omega_m$, $w_x$, $\xi_x$) and two-dimensional (2D) contours with $68\%$ C.L., $95\%$ C.L., and $99.7\%$ C.L.. From Table \ref{tab:results-mean-uxpos-3k} and Fig. \ref{fig:contour-uxpos-3k}, it is easy to see that the model parameter space is almost the same when the wavenumber $k\in(0.01,0.20)[hMpc^{-1}]$, particularly, the mean value of interaction rate is in the same order. This conclusion is in line with the analysis of the growth history in the right panel of Fig. \ref{fig:growth-kzf}. It means that, we could safely obtain the model parameter space for the coupled model $Q^{\mu}\parallel u^{\mu}_{(x)}$ from the redshift-space distortion measurement.

\begingroup
\squeezetable
\begin{center}
\begin{table}
\begin{tabular}{|cc|cc|cc|cc|}
\hline\hline Parameters & Priors & mean ($k=0.015$) & best-fit & mean ($k=0.065$) & best-fit & mean ($k=0.165$) & best-fit \\ \hline
$\Omega_bh^2$&[0.005,0.1]&
$0.0223_{-0.000244-0.000467}^{+0.000238+0.000477}$&$0.0224$&
$0.0223_{-0.000245-0.000493}^{+0.000265+0.000477}$&$0.0224$&
$0.0223_{-0.000245-0.000473}^{+0.000252+0.000499}$&$0.0225$
\\
$\Omega_ch^2$&[0.01,0.99]&
$0.114_{-0.00193-0.00432}^{+0.00224+0.00419}$&$0.115$&
$0.114_{-0.00171-0.00393}^{+0.00212+0.00366}$&$0.115$&
$0.114_{-0.00184-0.00419}^{+0.00221+0.00403}$&$0.115$
\\
$100\theta_{MC}$&[0.5,10]&
$1.0416_{-0.000569-0.00112}^{+0.000585+0.00118}$&$1.0414$&
$1.0416_{-0.000595-0.00110}^{+0.000572+0.00111}$&$1.0412$&
$1.0416_{-0.000583-0.00114}^{+0.000568+0.00116}$&$1.0417$
\\
$\tau$&[0.01,0.8]&
$0.0875_{-0.0138-0.0239}^{+0.0113+0.0252}$&$0.0964$&
$0.0887_{-0.0134-0.0236}^{+0.0119+0.0255}$&$0.0830$&
$0.0888_{-0.0121-0.0235}^{+0.0119+0.0252}$&$0.0890$
\\
$\xi_x$&[0,1]&
$0.00358_{-0.00358-0.00358}^{+0.000826+0.00609}$&$0.00149$&
$0.00328_{-0.00328-0.00328}^{+0.000736+0.00549}$&$0.00142$&
$0.00353_{-0.00353-0.00353}^{+0.000735+0.00587}$&$0.000923$
\\
$w_x$&[-1,0]&
$-0.970_{-0.0296-0.0296}^{+0.00718+0.0480}$&$-0.978$&
$-0.971_{-0.0292-0.0292}^{+0.00644+0.0443}$&$-0.994$&
$-0.970_{-0.0297-0.0297}^{+0.00654+0.0461}$&$-0.998$
\\
$n_s$&[0.5,1.5]&
$0.977_{-0.00562-0.0106}^{+0.00558+0.0110}$&$0.979$&
$0.977_{-0.00523-0.0106}^{+0.00539+0.0108}$&$0.978$&
$0.977_{-0.00547-0.0113}^{+0.00610+0.0112}$&$0.979$
\\
${\rm{ln}}(10^{10}A_s)$&[2.4,4]&
$3.0828_{-0.0258-0.0435}^{+0.0225+0.0482}$&$3.106$&
$3.0851_{-0.0256-0.0471}^{+0.0237+0.0494}$&$3.0729$&
$3.0853_{-0.0238-0.0449}^{+0.0230+0.0478}$&$3.0873$
\\
\hline
$\Omega_x$&$-$&
$0.706_{-0.00996-0.0206}^{+0.00997+0.0199}$&$0.702$&
$0.706_{-0.00953-0.0199}^{+0.00961+0.0180}$&$0.710$&
$0.706_{-0.00922-0.0204}^{+0.0104+0.0191}$&$0.712$
\\
$\Omega_m$&$-$&
$0.294_{-0.00997-0.0199}^{+0.00996+0.0206}$&$0.298$&
$0.294_{-0.00961-0.0179}^{+0.00953+0.0199}$&$0.290$&
$0.294_{-0.0104-0.0191}^{+0.00921+0.0204}$&$0.288$
\\
$\sigma_8$&$-$&
$0.805_{-0.0124-0.0244}^{+0.0122+0.0244}$&$0.815$&
$0.805_{-0.0122-0.0241}^{+0.0123+0.0240}$&$0.803$&
$0.806_{-0.0119-0.0253}^{+0.0128+0.0232}$&$0.810$
\\
$z_{re}$&$-$&
$10.689_{-1.124-1.967}^{+1.0158+2.0392}$&$11.485$&
$10.793_{-1.0337-2.0605}^{+1.0497+2.0563}$&$10.313$&
$10.804_{-1.0242-2.0596}^{+1.00808+2.0552}$&$10.805$
\\
$H_0$&$-$&
$68.317_{-0.862-1.781}^{+0.933+1.671}$&$68.191$&
$68.297_{-0.775-1.745}^{+0.966+1.571}$&$68.889$&
$68.281_{-0.804-1.754}^{+0.902+1.629}$&$69.246$
\\
${\rm{Age}}/{\rm{Gyr}}$&$-$&
$13.789_{-0.0370-0.0746}^{+0.0368+0.0736}$&$13.791$&
$13.790_{-0.0372-0.0732}^{+0.0377+0.0736}$&$13.784$&
$13.791_{-0.0376-0.0752}^{+0.0371+0.0749}$&$13.755$
\\
\hline\hline
\end{tabular}
\caption{The mean values with $1,2\sigma$ errors and the best-fit values of the parameters on the $\xi$wCDM model for several different scales ($k=0.015, 0.065$, and $0.165[hMpc^{-1}]$), where CMB from Planck + WMAP9, BAO, SNIa, and RSD data sets have been used.}
\label{tab:results-mean-uxpos-3k}
\end{table}
\end{center}
\endgroup

\begin{figure}[!htbp]
\includegraphics[width=20cm,height=15cm]{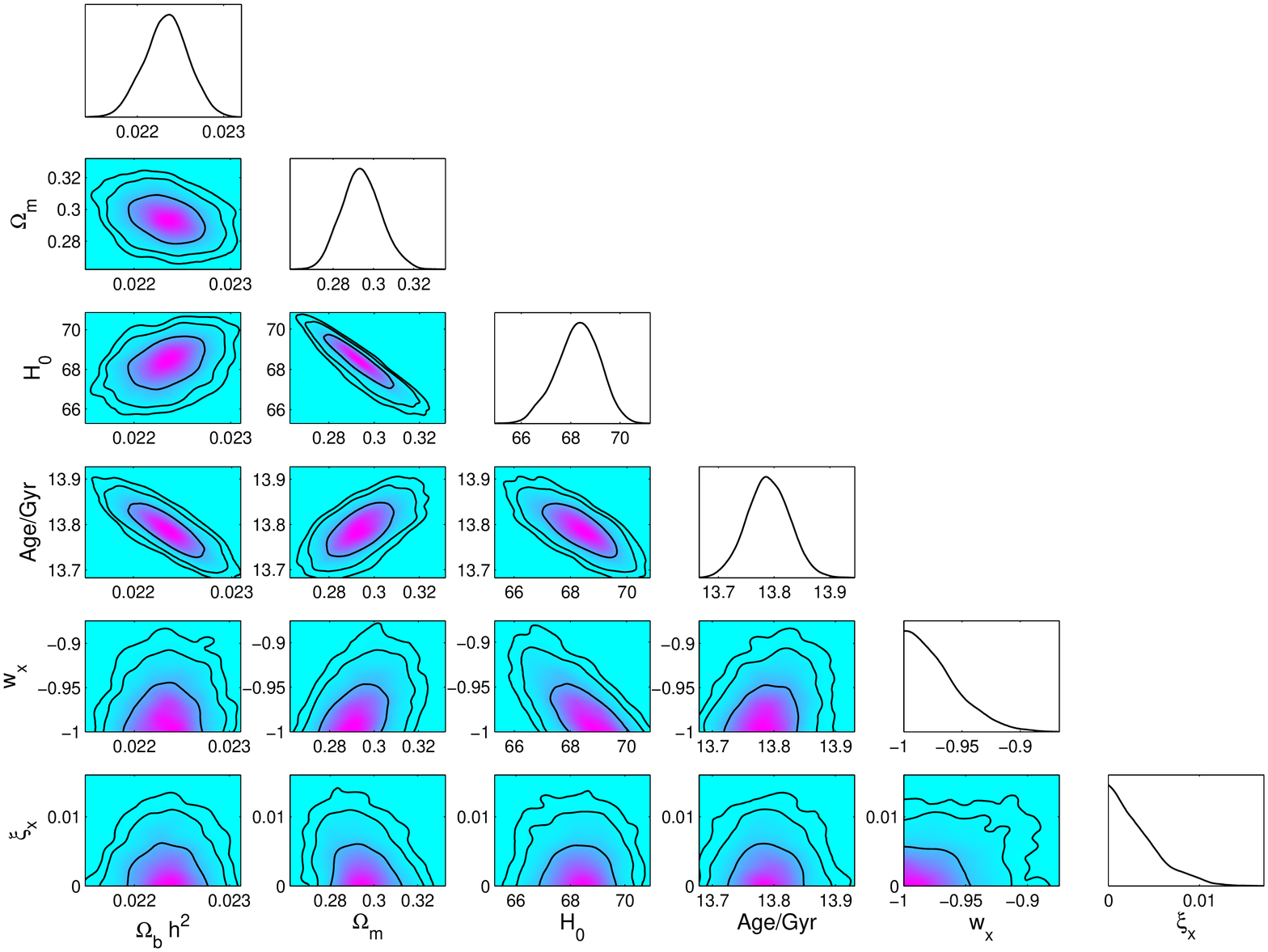}
  \caption{The 1D marginalized distributions on individual parameters and 2D contours between each other with 68\% C.L., 95 \% C.L., and 99.7\%, where CMB from Planck + WMAP9, BAO, SNIa, and RSD data sets have been used.}
  \label{fig:contour-uxpos}
\end{figure}

Then, choosing one of the constraint results ($k=0.065[hMpc^{-1}]$), we concretely show the mean values with $1,2,3\sigma$ errors and best-fit values of the parameters in Table \ref{tab:results-mean-uxpos}, and the one-dimensional (1D) marginalized distributions of the parameters and two-dimensional (2D) contours with $68\%$ C.L., $95\%$ C.L., and $99.7\%$ C.L. in Figs. \ref{fig:contour-uxpos}. In order to compare with the constraint without RSD data, we also constrain the $\xi$wCDM model without the $f\sigma_8(z)$ data set, the results are shown in the second and third columns of Table \ref{tab:results-mean-uxpos}. Obviously, the results with RSD test give a tighter constraint. Moreover, the interaction rate is more sensitive than the other parameters, which agrees with the discussion about modified growth of structure. Furthermore, based on the same observed data sets (CMB from Planck + WMAP9, BAO, SN and RSD), the $\xi$wCDM model has another two parameters $w_x$ and $\xi_x$ which give rise to the difference of the minimum $\chi^2$ with the $\Lambda$CDM model, $\Delta\chi^2_{min}=2.593$. Besides, we compare our constraint results of $Q^{\mu}\parallel u^{\mu}_{(x)}$ with ones of $Q^{\mu}\parallel u^{\mu}_{(c)}$ in Ref. \cite{ref:Yang2014-uc}.
%, the same coupling form $Q=3H\xi_x\rho_x$ was considered, the model has been tested by CMB from Planck + WMAP9, BAO, SNIa and RSD data. The results of Ref. \cite{ref:Yang2014-uc} showed the interaction rate in 3$\sigma$ regions: $\xi_x=0.00372_{-0.00372-0.00372-0.00372}^{+0.000768+0.00655+0.0102}$. They declared that the measurement of redshift-space distortion could rule out large interaction rate in 1$\sigma$ region. In this paper,
We obtain the compatible constraint results and the same conclusion with Ref. \cite{ref:Yang2014-uc}. From the fourth column of Table \ref{tab:results-mean-uxpos}, we find the recently cosmic observations indeed favor small interaction rate $\xi_x=0.00328_{-0.00328-0.00328-0.00328}^{+0.000736+0.00549+0.00816}$ after the RSD measurement is added. Moreover, in 1$\sigma$ region, the $f\sigma_8(z)$ test could rule out large interaction rate.

\begingroup
\squeezetable
\begin{center}
\begin{table}
\begin{tabular}{|c|cc|cc|cc|}
\hline\hline Parameters & mean ($\xi$wCDM without RSD) & best-fit & mean ($\xi$wCDM with RSD) & best-fit & mean ($\Lambda$CDM with RSD) & best-fit \\ \hline
$\Omega_bh^2$&
$0.0221_{-0.000247-0.000498-0.000627}^{+0.000252+0.000490+0.000624}$&$0.0220$&
$0.0223_{-0.000245-0.000493-0.000641}^{+0.000265+0.000477+0.000609}$&$0.0224$&
$0.0223_{-0.000245-0.000461-0.000602}^{+0.000245+0.000495+0.000642}$&$0.0225$
\\
$\Omega_ch^2$&
$0.0915_{-0.00985-0.0131-0.0146}^{+0.00462+0.0171+0.0227}$&$0.0892$&
$0.114_{-0.00171-0.00393-0.00521}^{+0.00212+0.00366+0.00476}$&$0.115$&
$0.116_{-0.00145-0.00282-0.00365}^{+0.00144+0.00286+0.00376}$&$0.115$
\\
$100\theta_{MC}$&
$1.0428_{-0.000677-0.00147-0.00197}^{+0.000830+0.00130+0.00169}$&$1.0427$&
$1.0416_{-0.000595-0.00110-0.00138}^{+0.000572+0.00111+0.00143}$&$1.0412$&
$1.0407_{-0.000551-0.00105-0.00138}^{+0.000543+0.00105+0.00143}$&$1.0408$
\\
$\tau$&
$0.0941_{-0.0133-0.0257-0.0338}^{+0.0136+0.0276+0.0353}$&$0.0807$&
$0.0887_{-0.0134-0.0236-0.0321}^{+0.0119+0.0255+0.0360}$&$0.0830$&
$0.0860_{-0.0128-0.0229-0.0293}^{+0.0117+0.0250+0.0325}$&$0.0788$
\\
$\xi_x$&
$0.0753_{-0.00483-0.0453-0.0654}^{+0.0247+0.0247+0.0247}$&$0.0847$&
$0.00328_{-0.00328-0.00328-0.00328}^{+0.000736+0.00549+0.00816}$&$0.00142$&
$---$&$-$
\\
$w_x$&
$-0.973_{-0.0270-0.0270-0.0270}^{+0.00477+0.0516+0.0821}$&$-0.994$&
$-0.971_{-0.0292-0.0292-0.0292}^{+0.00644+0.0443+0.0676}$&$-0.994$&
$---$&$-$
\\
$n_s$&
$0.972_{-0.00558-0.0113-0.0144}^{+0.00563+0.0115+0.0162}$&$0.970$&
$0.977_{-0.00523-0.0106-0.0133}^{+0.00539+0.0108+0.0142}$&$0.978$&
$0.969_{-0.00542-0.0109-0.0148}^{+0.00538+0.0107+0.0146}$&$0.972$
\\
${\rm{ln}}(10^{10}A_s)$&
$3.103_{-0.0260-0.0506-0.0642}^{+0.0268+0.0539+0.0695}$&$3.0772$&
$3.0851_{-0.0256-0.0471-0.0630}^{+0.0237+0.0494+0.0676}$&$3.0729$&
$3.0719_{-0.0232-0.0444-0.0565}^{+0.0232+0.0488+0.0630}$&$3.0559$
\\
\hline
$\Omega_x$&
$0.762_{-0.0139-0.0442-0.0586}^{+0.0253+0.0355+0.0406}$&$0.770$&
$0.706_{-0.00953-0.0199-0.0253}^{+0.00961+0.0180+0.0232}$&$0.710$&
$0.710_{-0.00819-0.0167-0.0224}^{+0.00815+0.0158+0.0120}$&$0.713$
\\
$\Omega_m$&
$0.238_{-0.0253-0.0355-0.0406}^{+0.0139+0.0442+0.0586}$&$0.230$&
$0.294_{-0.00961-0.0179-0.0232}^{+0.00953+0.0199+0.0253}$&$0.290$&
$0.290_{-0.00815-0.0158-0.0199}^{+0.00819+0.0167+0.0224}$&$0.287$
\\
$\sigma_8$&
$---$&$-$&
$0.805_{-0.0122-0.0241-0.0350}^{+0.0123+0.0240+0.0319}$&$0.803$&
$0.810_{-0.0109-0.0190-0.0249}^{+0.00992+0.0201+0.0263}$&$0.802$
\\
$z_{re}$&
$11.375_{-1.104-2.243-2.978}^{+1.149+2.239+2.860}$&$10.289$&
$10.793_{-1.0337-2.0605-2.896}^{+1.0497+2.0563+2.869}$&$10.313$&
$10.570_{-1.0183-2.0293-2.670}^{+1.0229+2.0821+2.635}$&$9.904$
\\
$H_0$&
$69.344_{-0.927-1.834-2.449}^{+0.932+1.851+2.174}$&$69.763$&
$68.297_{-0.775-1.745-2.264}^{+0.966+1.571+1.992}$&$68.889$&
$69.130_{-0.665-1.315-1.745}^{+0.677+1.336+1.692}$&$69.456$
\\
${\rm{Age}}/{\rm{Gyr}}$&
$13.784_{-0.0369-0.0728-0.0891}^{+0.0367+0.0701+0.0892}$&$13.793$&
$13.790_{-0.0372-0.0732-0.0927}^{+0.0377+0.0736+0.100}$&$13.784$&
$13.756_{-0.0344-0.0709-0.0932}^{+0.0376+0.0683+0.0897}$&$13.737$
\\
\hline\hline
\end{tabular}
\caption{The mean values with $1,2,3\sigma$ errors and the best-fit values of the parameters for the $\xi$wCDM model ($k=0.065[hMpc^{-1}]$) and the $\Lambda$CDM model, where CMB from Planck + WMAP9, BAO, SNIa, with or without RSD data sets have been used.}
\label{tab:results-mean-uxpos}
\end{table}
\end{center}
\endgroup

%\begin{figure}[!htbp]
%\includegraphics[width=20cm,height=15cm]{contour-uxpos-red.eps}
%  \caption{The 1D marginalized distributions on individual parameters and 2D contours between each other with 68\% C.L., 95 \% C.L., and 99.7\%, where CMB from Planck + WMAP9, BAO, SNIa, and RSD data sets have been used.}
%  \label{fig:contour-uxpos-red}
%\end{figure}

\section{SUMMARY}

Based on the large scale stability of the perturbations, we considered the coupling form $\bar{Q}=3H\xi_x\bar{\rho}_x$ as our research emphasis. For the dark energy model, we took dark energy as a fluid with a constant equation of state parameter $w_x$. The coupling between the dark components gave rise to the deviation from the uncoupled model. In the evolution of background equations, the effective EoS of dark energy was changed. As for the cosmological perturbations, we derived the evolution equations of the density and velocity perturbations for the $\xi$wCDM model by combining the background energy transfer with vanishing momentum transfer potential in the rest frame of dark energy. By changing the interaction rate $\xi_x$ and fixing the other parameters, we have gained the insight into the physical implications of the coupling between the dark components, which mainly included the evolutions of CMB temperature power spectra, matter power spectra, and $f\sigma_8(z)$.
From the comparison between the two coupled models with $b=0$ and $b=1$, it turned out that there actually were differences on the modified growth history between the two models, however, it was also clear that it would be vary hard even in the future to distinguish them.
Moreover, the figures told us that $f\sigma_8(z)$ evolution could significantly break the degeneracy between the $\xi$wCDM model and $w$CDM model, but the power spectra could not make it.
Both the theoretical analysis on the modified growth and the constraint results for several different scales told us that we could safely obtain the model parameter space with the redshift-space distortion measurement.
Then, using CMB from Planck + WMAP9, BAO, SNIa, and the RSD measurement, we conducted a full MCMC likelihood analysis for the coupled model. The joint data sets are able to estimate the parameter space to high precision and evidently tighten the constraints than the case without $f\sigma_8(z)$ test. The results showed the interaction rate in 3$\sigma$ regions: $\xi_x=0.00328_{-0.00328-0.00328-0.00328}^{+0.000736+0.00549+0.00816}$. It meant that the currently available cosmic observations favored a small interaction rate which is up to the order of $10^{-2}$, at the same time, the $f\sigma_8(z)$ test could rule out the large interaction rate in 1$\sigma$ region.

In the future work, when the expansion rate of the Universe is perturbed \cite{ref:Gavela2010}, we try to constrain the coupled model by combining the RSD measurement with the geometry test. It is worthwhile to anticipate that the large scale structure measurement will help to significantly tighten the constraints for some other cosmological models and give the constructive suggestions for theoretical research.

\acknowledgements{L. Xu's work is supported in part by NSFC under the Grants No. 11275035 and "the Fundamental Research Funds for the Central Universities" under the Grants No. DUT13LK01.}

\end{document}